\documentclass[iop]{emulateapj}

\def\lapp{\ifmmode\stackrel{<}{_{\sim}}\else$\stackrel{<}{_{\sim}}$\fi}
\def\gapp{\ifmmode\stackrel{>}{_{\sim}}\else$\stackrel{>}{_{\sim}}$\fi}
\usepackage{multirow}
\usepackage{color}
\usepackage{amsmath}
\usepackage{soul}

\begin{document}

\title{Broadband X-ray Properties of the Gamma-ray Binary 1FGL~J1018.6$-$5856}

\author{
Hongjun~An\altaffilmark{1,2},
Eric~Bellm\altaffilmark{3},
Varun~Bhalerao\altaffilmark{4},
Steven~E.~Boggs\altaffilmark{5},
Finn~E.~Christensen\altaffilmark{6},
William~W.~Craig\altaffilmark{5,7},
Felix~Fuerst\altaffilmark{3},
Charles~J.~Hailey\altaffilmark{8},
Fiona~A.~Harrison\altaffilmark{3},
Victoria~M.~Kaspi\altaffilmark{2,9},
Lorenzo~Natalucci\altaffilmark{10},
Daniel~Stern\altaffilmark{11},
John~A.~Tomsick\altaffilmark{5},
and William~W.~Zhang\altaffilmark{12}
\\
{\small $^1$Department of Physics/KIPAC, Stanford University, Stanford, CA 94305-4060, USA}\\
{\small $^2$Department of Physics, McGill University, Rutherford Physics Building,
3600 University Street, Montreal, QC H3A 2T8, Canada} \\
{\small $^3$Cahill Center for Astronomy and Astrophysics,
California Institute of Technology, Pasadena, CA 91125, USA} \\
{\small $^4$Inter University Center for Astronomy and Astrophysics,
Post Bag 4, Ganeshkhind, Pune 411007, India}\\
{\small $^5$Space Sciences Laboratory, University of California, Berkeley, CA 94720, USA}\\
{\small $^6$DTU Space, National Space Institute, Technical University of Denmark,
Elektrovej 327, DK-2800 Lyngby, Denmark}\\
{\small $^7$Lawrence Livermore National Laboratory, Livermore, CA 94550, USA}\\
{\small $^8$Columbia Astrophysics Laboratory, Columbia University, New York NY 10027, USA}\\
{\small $^9$McGill Space Institute, 3550 University Street, Montreal, QC H3A 2T8, Canada} \\
{\small $^{10}$Istituto Nazionale di Astrofisica, INAF–IAPS, via del Fosso del Cavaliere,
I-00133 Roma, Italy}\\
{\small $^{11}$Jet Propulsion Laboratory, California Institute of Technology,
Pasadena, CA 91109, USA}\\
{\small $^{12}$Goddard Space Flight Center, Greenbelt, MD 20771, USA}\\
}

\begin{abstract}
We report on {\it NuSTAR}, {\it XMM-Newton} and {\it Swift} observations of
the gamma-ray binary 1FGL~J1018.6$-$5856.
We measure the orbital period to be $16.544\pm0.008$ days
using {\it Swift} data spanning 1900 days.
The orbital period is different from the 2011 gamma-ray measurement which was used in the
previous X-ray study of \citet{adk+13} using $\sim$400\,days of {\it Swift} data,
but is consistent with
a new gamma-ray solution reported in 2014. The light curve folded on the
new period is qualitatively similar to that reported previously, having a
spike at phase 0 and broad sinusoidal modulation.
The X-ray flux enhancement at phase 0
occurs more regularly in time than was previously suggested. A spiky structure
at this phase seems to be a persistent feature, although there is
some variability. Furthermore, we find that the source flux
clearly correlates with the spectral hardness throughout all orbital phases,
and that the broadband X-ray spectra measured with {\it NuSTAR},
{\it XMM-Newton}, and {\it Swift} are well fit with an unbroken power-law model.
This spectrum suggests that the system may not be accretion-powered.
\end{abstract}


\section{Introduction}
Gamma-ray binaries are systems composed of a massive star and a
compact object and from which persistent
GeV and/or TeV gamma-ray emission is detected and
dominates the overall non-thermal spectrum.
They emit across the electromagnetic spectrum
from the radio to TeV gamma ray \citep[see][for a review]{m12}.
There are only five
gamma-ray binaries known to date \citep[][]{d13},
and only for one source has the compact object been identified
\citep[PSR~B1259$-$63;][]{jml+92}.

Since most of the energy output of a gamma-ray binary
is in the gamma-ray band, current theoretical studies focus on
explaining the high energy emission properties.
The gamma-ray emission models can be categorized
into two 
classes: microquasar models \citep[e.g.,][]{rtk+03, bp04} and
pulsar models \citep[e.g.,][]{tak94,st08}.
In the microquasar model,
relativistic electrons in a jet generated close to the compact object Compton-upscatter
the synchrotron emission of the jet itself and/or
the stellar UV photons \citep[e.g.,][]{krm02, bp04}, or
relativistic hadrons collide with background nuclei creating pions that
decay \citep[e.g.,][]{rtk+03}, producing gamma rays.
In the pulsar model, pulsar wind particles are accelerated in the pulsar wind/stellar wind shock,
and Compton-upscatter stellar photons to produce the observed gamma rays
\citep[e.g.,][]{tak94, ta97, d06, st08}.

Non-thermal X-ray emission in gamma-ray binaries is thought to be produced
by the electrons which are accelerated in the pulsar wind/stellar wind shock
\citep[e.g.][]{ta97,d06}
or in relativistic jets formed close to the compact object \citep[e.g.,][]{bk09}.
The models predict varying X-ray fluxes and spectra
depending on the properties of the shock, which are determined
by the thrust of the winds
and the orbital geometry of the binary system \citep[e.g.,][]{ktn+95},
or on the jet dynamics and cooling timescale \citep[e.g.,][]{dch10,bk09}.
Hence, X-ray measurements can be used for
constraining the orbital parameters and understanding the nature of
the physical processes in gamma-ray binaries
\citep[see also][]{chw06, tku+09, ton+12}.

The gamma-ray binary 1FGL~J1018.6$-$5856 was discovered with {\it Fermi} in 2011.
\citet{fermi12} found modulation in the radio to gamma-ray bands with a period of
$16.58 \pm 0.02$\,days, identifying the source as a gamma-ray binary. They further identified
the companion star to be an O6V((f)) star. Soon after the discovery,
subsequent broadband studies
were carried out \citep{ltc+11,hess12,adk+13} in order to better characterize the
source properties, but in no case were they able to identify the nature of the compact object.

X-ray properties of the gamma-ray binary 1FGL~J1018.6$-$5856 were measured in detail
with {\it Swift}. \citet{adk+13}
showed that the X-ray flux peak seen at phase 0
(gamma-ray maximum) by \citet{fermi12} seems not to be a persistent feature
and instead shows a relatively large orbit-to-orbit variation.
Furthermore, \citet{adk+13} found evidence of a correlation
between flux and spectral hardness in the X-ray band.

Recently, \citet{ccc+14} refined the gamma-ray period using {\it Fermi} observations
with a longer baseline, and found the period to be $16.531\pm0.006$\,days. Since
this is slightly different from the value ($16.58\pm0.02$\,days) used
for the previous X-ray study carried out by \citet{adk+13},
the X-ray results need to be refined using the new gamma-ray period.
The baseline of the X-ray observations is long (5 years), and
thus phases of later observations may change significantly.

Important questions to be addressed for gamma-ray binaries are:
what is the nature of
the compact object \citep[known only for PSR~B1259$-$63,][]{jml+92},
and what is the physical emission mechanism.
If the source is powered by accretion, a complex continuum spectrum
is expected whether the compact object is a neutron star or a black hole.
Hence, accurate measurement of the spectrum will help us identify
the compact object. Furthermore,
searching for a spectral turn-over in the hard X-ray band
\citep[e.g.,][]{gjk+99, chr+02}
and/or spectral lines often seen in high-mass X-ray
binaries (HMXBs) may also provide clues about the emission mechanism
of the source.

In this paper, we measure X-ray properties of the gamma-ray binary
1FGL~J1018.6$-$5856 more
accurately than before using new observations taken with {\it NuSTAR},
{\it Swift} and with archival {\it XMM-Newton} observations.
In Section~\ref{sec:sec1}, we describe the observations
we used in this paper. We show data analysis and the results in Section~\ref{sec:ana}.
We then discuss our findings in Section~\ref{sec:disc}, and conclude in Section~\ref{sec:concl}.

\section{Observations}
\label{sec:sec1}

\newcommand{\marka}{\tablenotemark{a}}
\newcommand{\markb}{\tablenotemark{b}}
\newcommand{\markc}{\tablenotemark{c}}
\begin{table*}[t]
\vspace{-0.0in}
\begin{center}
\caption{Summary of observations used in this work and results of spectral analysis
\label{ta:ta1}}
\vspace{-0.05in}
\scriptsize{
\begin{tabular}{ccccccccc} \hline\hline
Observatory     &Obs. ID        & Date     & Exposure  & $\phi$ & $N_{\rm H}$ & $\Gamma$ & $F_{\rm 3-10\ keV}$\marka  & Mode   \\ 
                &               & (MJD)    & (ks)      & & ($10^{22}\rm cm^{-2}$) &   & ($\rm \ erg\ s^{-1}\ cm^{-2}) $ &     \\ \hline 
{\em XMM}& 0604700101 & 55066  & 20/12  & 0.6 & 0.65(5) & 1.65(7) & $5.1\pm0.2\times 10^{-13}$ & FW/FW\markb  \\ 
{\em XMM}& 0694390101 & 56302  & 104/73 & 0.3 & 0.72(2) & 1.57(2) & $1.09\pm0.01\times 10^{-12}$ & FW/SW\markb  \\ 
\multirow{2}{*}{\em Swift}     & 00031912001-- & 55103-- & 
\multirow{2}{*}{169} & \multirow{2}{*}{0.0--1.0} & \multirow{2}{*}{0.72\markc} & \multirow{2}{*}{1.2--1.8} & \multirow{2}{*}{$0.34-1.2\times 10^{-12}$} & \multirow{2}{*}{PC} \\ 
                               & 00090191001 & 56992 &  & & & &  \\ 
{\em NuSTAR}    & 30002020002   &  56812  & 22 & 0.2 & 0.72\markc & 1.67(10) & $5.7\pm0.5\times 10^{-13}$ &  $\cdots$  \\  
{\em NuSTAR}    & 30002020004   &  56862  & 23 & 0.2 & 0.72\markc & 1.77(10) & $6.1\pm0.5\times 10^{-13}$ &  $\cdots$  \\ 
{\em NuSTAR}    & 30002020006   &  56911  & 25 & 0.2 & 0.72\markc & 1.64(8)  & $7.8\pm0.5\times 10^{-13}$ &  $\cdots$  \\ 
{\em NuSTAR}    & 30002020008   &  56992  & 21 & 0.0 & 0.72\markc & 1.41(7)  & $1.11\pm0.06\times 10^{-12}$ &  $\cdots$  \\  \hline
\end{tabular}}
\end{center}
$^{\rm a}${ Absorption-corrected flux.}\\
$^{\rm b}${ For MOS1,2/PN. FW: Full window. SW: small window.}\\
$^{\rm c}${ $N_{\rm H}$ was frozen for the {\it Swift} and {\it NuSTAR} data fit.}\\
\vspace{-1.0 mm}
\end{table*}

We observed the gamma-ray binary 1FGL~J1018.6$-$5856
with {\it NuSTAR} \citep[][]{hcc+13} four times between 2014 June 4 and December 1
with exposures of $\sim$20\,ks for each observation.
The total exposure was 90 ks.
Soft X-ray band below and overlapping with the {\it NuSTAR} band (3--79\,keV)
was covered with {\it Swift} observations
and two archival {\em XMM-Newton} observations (see Table~\ref{ta:ta1}).
The total exposure of the 71 {\it Swift} observations was 169\,ks, and each
exposure was relatively short.

The {\it NuSTAR} observations were processed with the standard pipeline
tools {\tt nupipeline} and {\tt nuproducts} of {\tt nustardas}
1.4.1 integrated in HEASOFT 6.16.
We used {\it NuSTAR} CALDB version 20140414 and applied the standard
filters.\footnote{See http://heasarc.gsfc.nasa.gov/docs/nustar/analysis/nustar\\\_swguide.pdf for more details}
In order to process the {\it Swift} data, we used the {\ttfamily xrtpipeline} tool along with HEASARC
remote CALDB\footnote{http://heasarc.nasa.gov/docs/heasarc/caldb/caldb\_remote\_ac\\cess.html}
and standard filters \citep{cps+05}.
Note that the source was not clearly detected in some {\it Swift} observations,
and that the {\it Swift} observations taken until MJD~55984
were reported previously \citep[][]{fermi12, adk+13}.
The {\it XMM-Newton} data were processed with {\ttfamily epproc} and {\ttfamily emproc} in
Science Analysis System (SAS) 14.0.0\footnote{http://xmm.esac.esa.int/sas/}
using standard filters.

\section{Data Analysis and Results}
\label{sec:ana}

\subsection{Timing Analysis}
\label{timingana}
Detection of pulsations in gamma-ray binaries can be difficult for
several reasons, such as the possibilities of an unfavorable
emission geometry, absorption of soft X-rays by the wind, or a large background
due to non-thermal unpulsed emission.
Even in a favorable situation where the above effects
are minimal, the Doppler effect due to binary motion can blur the pulse signal
if the orbit is tight.
For 1FGL~J1018.6$-$5856, \citet{adk+13} showed that the Doppler broadening is
not a concern for a 20-ks observation assuming a circular orbit
with an inclination of 30$^\circ$. We therefore attempt
to search for the pulsation.
Event arrival times measured at the spacecraft were transformed into those
at the solar system barycenter
with {\ttfamily barycorr} for the {\it NuSTAR} and {\ttfamily barycen}
for the {\em XMM-Newton} data. We did not search the {\it Swift} data because of
the paucity of counts in individual {\it Swift} observations.

For the {\it NuSTAR} data, we produced an event list for each observation
in the 3--20\,keV band using a circular aperture with $R=30''$.
We performed the timing analysis with the data
from each {\it NuSTAR} focal plane module\footnote{{\it NuSTAR} has
two focal plane modules, FPMA and FPMB.}
as well as with the combined dataset.
Above 20\,keV background dominates, and hence we adopt that as the high end of our band.
Note that the results below do not
depend strongly on the exact energy range or the aperture size. We folded
the event time series to test periods between $10^{-4}-10^{3}$ s,
and calculated $Z_1^2$ \citep{bbb+83}. We find that $Z_1^2$ is fairly large
for some test periods. 
However, we find that the large $Z_1^2$ seen in one observation is
not reproduced in the others.  We further verified that the large $Z_1^2$
values are not significant. Note that the measured $Z_1^2$ distribution
does not follow a $\chi^2$ distribution, but has a long tail, and thus
we used a functional distribution obtained by fitting the measured $Z_1^2$
distribution in order to estimate the significance.
We performed the same study for the {\it XMM-Newton}/PN data in the 0.5--2\,keV
and 0.5--10\,keV bands, and did not find any significant pulsations.
Assuming the pulse profile is a sine function with a period in the range of 0.1--1\,s,
we estimate the 90\% upper limit for the pulse fractions to be 47\% and 6\% in
the 3--20 keV and 0.5--10 keV bands, respectively.

\begin{figure}
\centering
\vspace{-0.1in}
\hspace{-0.45in}
\includegraphics[width=3.6 in]{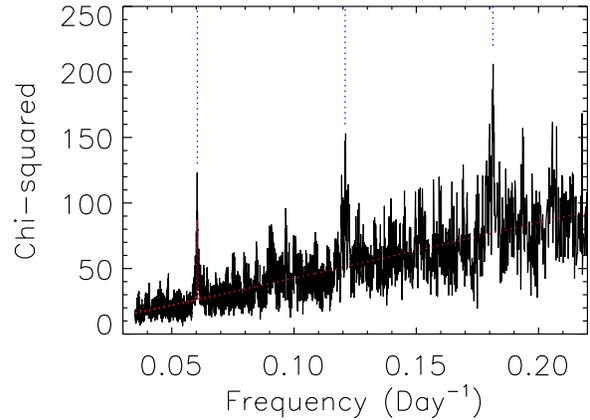} 
\vspace{-0.1in}
\figcaption{Chi-squared vs. frequency for the 0.5--10\,keV {\em Swift } X-ray data.
The search step is 0.01\,day and, the number of phase bins is 16.
Frequencies for the first three harmonics are denoted as blue dotted lines and
the best-fit function is shown in red.
\label{fig:fig1}
}
\end{figure}

Next, we refine the X-ray measurement of the orbital period
by using a longer baseline using the {\it Swift} data over a longer time period
than the previous work.
Note that we did not use the data taken with {\it XMM-Newton} or {\it NuSTAR}
because their count rate measurements cannot be directly compared to those of {\it Swift}.
As was done by \citet{adk+13}, we use epoch folding \citep{l87} because of the unequal
exposures of the observations. In the {\it Swift} observations,
we extracted source and background events in the 0.5--10\,keV band within
a $30''$ radius circle, and an annular region with inner radius 50$''$
and outer radius 100$''$, respectively.
We then folded the event time series at test
periods around $P_{\rm orb}=16.531$\,days \citep[][]{ccc+14},
producing a light curve with 16 bins. We used the same epoch for phase 0 as
that used in the previous studies \citep[][]{fermi12, adk+13}.
We calculated $\chi^2$ of the light curve for each trial period, and followed the fitting
technique as described in \citet{l87}. Note, however, that we
modeled the underlying continuum using a power-law function instead of
the constant model employed by \citet{l87},
because the $\chi^2$ of the folded light curve
is rising towards short periods (see Figure~\ref{fig:fig1}). The best-fit continuum
model is $\chi^2_{\rm cont} \sim P^{-1}$; the exact value of the power-law index
varies between 0.9 and 1.1 depending on the fit range.

\begin{figure*}[ht]
\centering
\vspace{8mm}
\begin{tabular}{ccc}
\hspace{-8.00mm}
\vspace{-5.00mm}
\includegraphics[width=2.7 in]{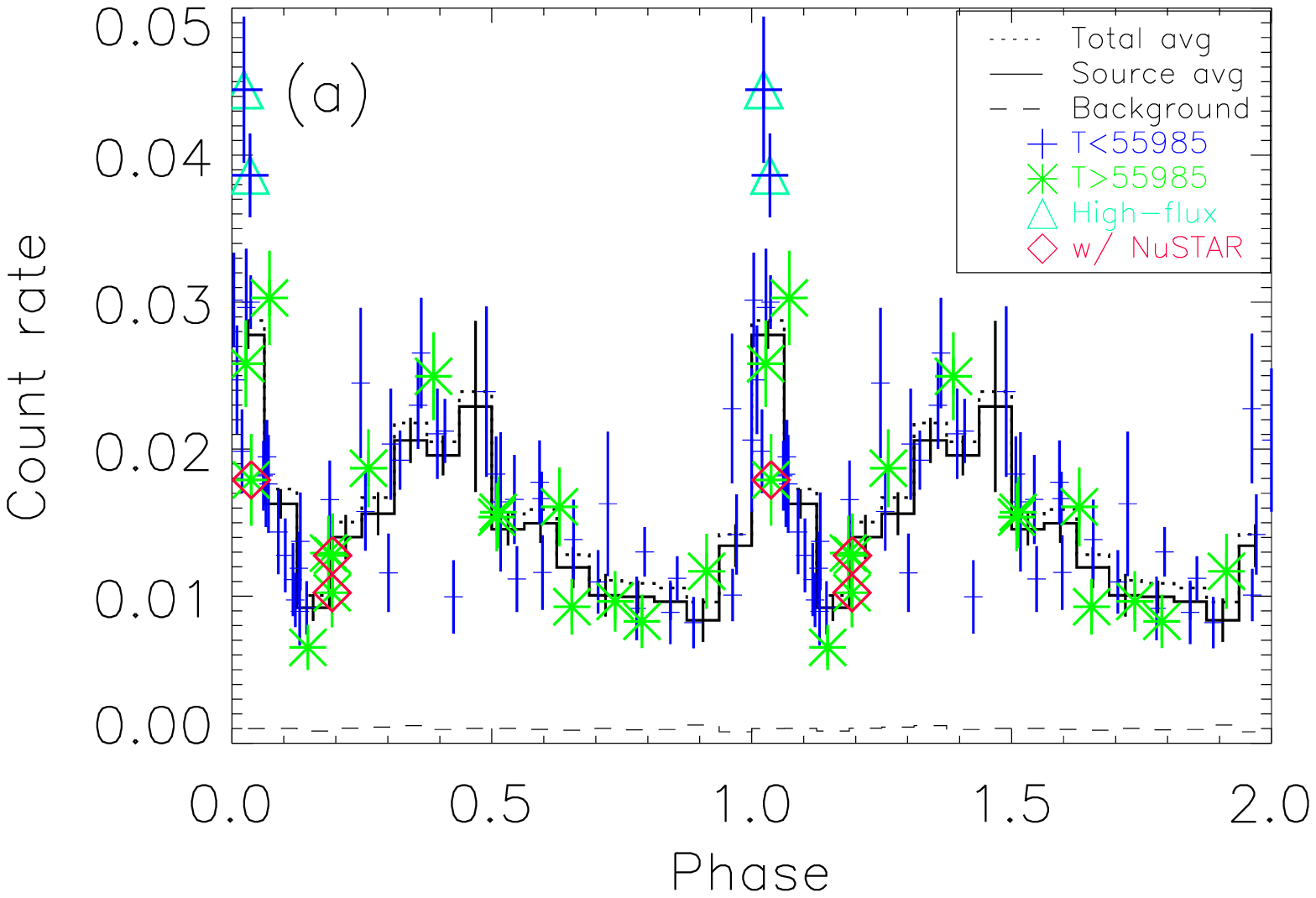} &
\hspace{-13.00mm}
\includegraphics[width=2.7 in]{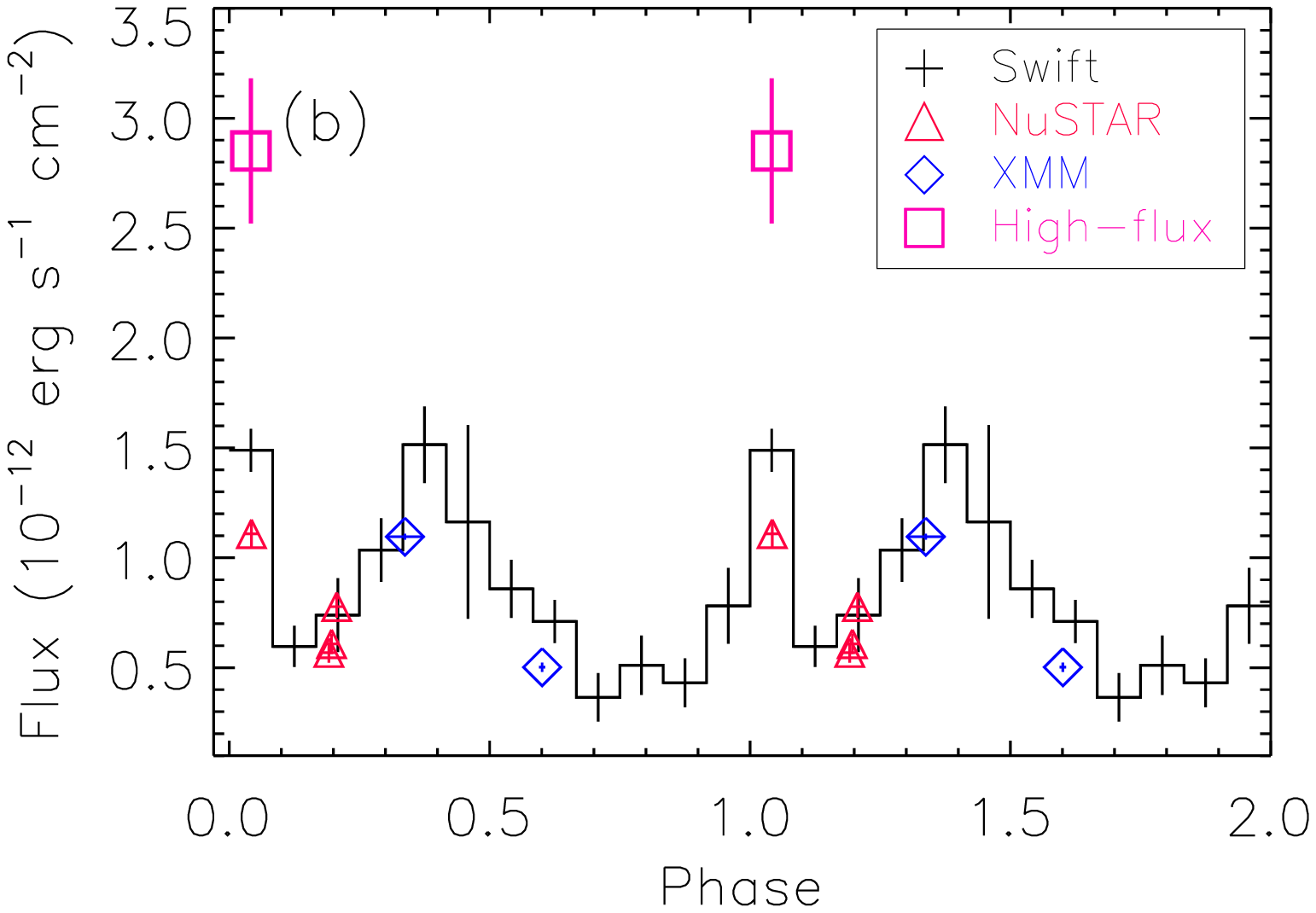} &
\hspace{-13.00mm}
\includegraphics[width=2.7 in]{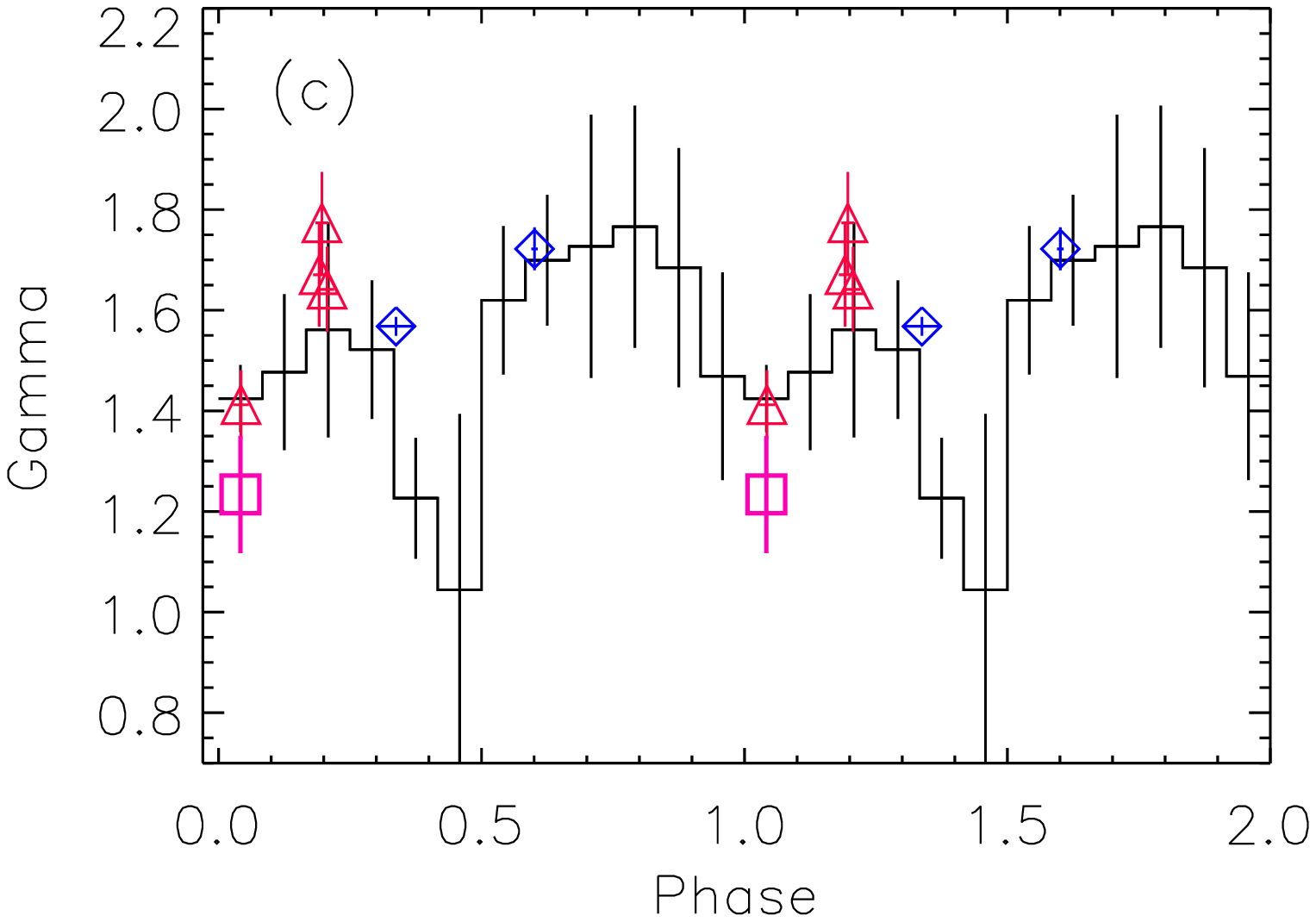} \\
\end{tabular}
\figcaption{X-ray light curve and the best-fit spectral parameters for a power-law model.
Only {\it Swift} data are shown in panel (a)  and measurements made with all three
instruments are shown in panels (b) and (c).
{\it a}) 0.5--10\,keV light curve as measured with {\it Swift}.
Black dotted line shows the average count rate,
black solid line is for the average source count rate, and the black dashed line shows the average
background count rate. Blue data points are measurements reported by
\citep[before 55985 MJD][]{adk+13}, and green data points are new measurements
\citep[][and this work]{ccc+14}. Cyan triangle denotes the ``high-flux state'',
two observations which had significantly
larger count rates than the others at the same phase, and red diamonds are for
time periods in which {\it NuSTAR} observations were made.
{\em b}) 3--10\,keV flux corrected for interstellar absorption.
Data points for {\it Swift}, {\it NuSTAR},
and the {\it XMM-Newton} measurements are denoted in cross, triangle, and diamond, respectively.
{\em c}) the best-fit photon index.
Same symbols as in (b) are used in (c).
\label{fig:fig2}
}
\end{figure*}

We find that the best-fit orbital period varies between 16.538 and 16.55 days,
depending on the number of bins, and the search step or search range.
We varied the number of bins between 8 and 18 to ensure that the light curve
is resolved and the $\chi^2$ statistic is applicable, and the search step between 0.001
and 0.005 days, smaller than the uncertainty in the $P_{\rm orb}$ measurement
($\Delta P_{\rm orb}\lapp 0.01$\,days). The search range was varied between $\pm 1$\,day
and $\pm15$\, days. We find that the variations
are within 1$\sigma$ of the measurements.
The resulting period is $P_{\rm orb}=16.544\pm0.008$ days.
We show the folded light curve in Figure~\ref{fig:fig2}a.

We also used the second harmonic to measure the orbital period, since it can
be measured with better precision, and found that the measurement is more
stable, varying only $6\times 10^{-4}$\,days as a function of the number of bins, search step or
the search range. The result is $P_{\rm orb}=16.543\pm0.004$ days.
Our result is consistent with the {\em Fermi} measured value
\citep[$P_{\rm orb}=16.531\pm0.006\rm \ days$,][]{ccc+14} with
a null hypothesis probability $p=0.09$.

\subsection{Spectral Analysis}
\label{spectrumana}

\begin{figure*}[ht]
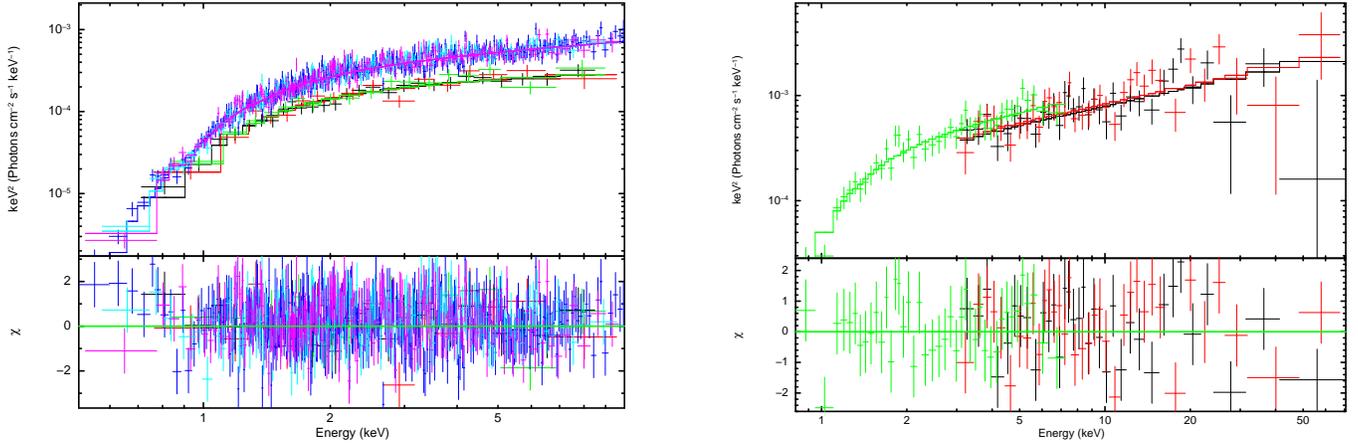

\centering
\hspace{-2 mm}
\begin{tabular}{cc}
\hspace{-6 mm}
\vspace{3 mm}
\includegraphics[width=2.3 in, angle=-90]{./fig3a.eps} &
\hspace{5 mm}
\includegraphics[width=2.3 in, angle=-90]{./fig3b_new.eps} \\
\hspace{-3 mm}
\end{tabular}
\figcaption{Broadband X-ray spectra obtained with {\it XMM-Newton},
{\it NuSTAR} and {\it Swift}.
{\it Left}: {\it XMM-Newton} spectra obtained at phases 0.3 and 0.6.
The harder and brighter spectra for phase 0.3 measured with PN, MOS1 and MOS2
are colored in blue, cyan and magenta, respectively and
the spectra for phase 0.6 are colored in black, red and green
(see Figures~\ref{fig:fig2}b and c).
{\it Right}: {\it NuSTAR} and {\it Swift} spectra for phase 0 without the high-flux state.
Note that the {\it Swift} spectrum (green) is obtained by combining {\it Swift} observations
taken at phase 0 excluding the high-flux state. Black and red data points show the
{\it NuSTAR} spectra measured with FPMA and FPMB, respectively.
The best-fit power-law model is shown in solid lines in each panel.
\vspace{2 mm}
\label{fig:fig3}
}
\hspace{1mm}
\vspace{-2mm}
\end{figure*}

Although $N_{\rm H}$ towards the source has been measured previously,
the uncertainty was relatively large.
Since there is one more {\it XMM-Newton} observation
(Obs. ID 0694390101, Table~\ref{ta:ta1}) taken after
the previous X-ray study \citep[][]{adk+13},
we can determine $N_{\rm H}$ more precisely using the {\it XMM-Newton} observations.
We extracted the source spectrum from a circle with $R=16''$ (Obs. ID 0604700101) or
$R=24''$ (Obs. ID 0694390101), and background spectra from a circle with $R=32''$
in a source-free region
$\sim$200$''$ vertically upwards along the detector column from the source.
Note that we used different source extraction regions because of differences
in exposure times.

Since it has been suggested that the spectral hardness varies orbitally,
we used different spectral slopes for observations taken at different phases
(Figure~\ref{fig:fig2}). Thus, we fit the two {\it XMM-Newton} spectra separately
allowing all the fit parameters to vary.
We grouped the spectra to have 20 counts per bin,
and fit them with an absorbed power-law model with
the {\tt angr} abundance in {\tt XSPEC} \citep[][]{ag89}
using $\chi^2$ statistics or {\it l} statistics \citep[][]{l92}.
The two methods yield consistent results. The best-fit $N_{\rm H}$ values
for the observations are statistically consistent with each other (Table~\ref{ta:ta1}).
Best-fit $N_{\rm H}$ values obtained with a different abundance model
\citep[{\tt wilm} in {\tt XSPEC};][]{wam00}
for the two spectra are still consistent with each other.
Therefore, we use a common $N_{\rm H}$ value and
find that a power-law model successfully explains the data (Figure~\ref{fig:fig3}, left).
The best-fit value is $N_{\rm H}=7.2\pm0.2\times10^{21}\rm \ cm^{-2}$,
and we use this value throughout this paper.
Note that using $N_{\rm H}=7.2\pm0.2\times10^{21}\rm \ cm^{-2}$ does not change
the other spectral parameters in Table~\ref{ta:ta1} significantly.
We find that using the {\tt wilm} abundance model
changes the best-fit $N_{\rm H}$ values
(to $0.93\pm0.08\times10^{22}\rm \ cm^{-2}$, $1.03\pm0.02\times10^{22}\rm \ cm^{-2}$,
and $1.02\pm0.02\times10^{22}\rm \ cm^{-2}$ for Obs. IDs 0604700101, 0694390101 and combined, respectively),
but the other spectral parameters do not change significantly.
We note that the source count rates were less than 0.03--0.08 cps for MOS1/2, and
0.1--0.3 cps for PN, and hence pile-up is not a
concern.\footnote{http://xmm2.esac.esa.int/docs/documents/CAL-TN-0200-1-0.pdf}

For the {\it NuSTAR} data, we extracted source and background events
from circular regions
with $R=30''$ and $R=45''$, respectively. Backgrounds were extracted in the same detector
chip as the source, offset $\sim$4$'$ from the source region. The source was
detected above the background up to 20--30\,keV. We grouped the spectra to have a minimum of 20
counts per spectral bin, and used $\chi^2$ statistics and {\it l} statistics;
they provide consistent results.

We jointly fit the data with a power-law model having different photon indexes for
different orbital phases, and found that the best-fit parameters are
$\Gamma=1.69\pm0.05$ and $F_{\rm 3-10\ keV}=5.7\pm0.4 \times 10^{-13}\ \rm erg\ cm^{-2}\ s^{-1}$
for $\phi=0.2$, and
$\Gamma=1.41\pm0.07$ and $F_{\rm 3-10\ keV}=1.11\pm0.06 \times 10^{-12}\ \rm erg\ cm^{-2}\ s^{-1}$
for $\phi=0$.
We find that a power-law model with a constant photon index across orbital phases
can also explain the data if we let the flux vary between phases. However,
a model with separate spectral indices for different phases
provides a significantly better fit than one with constant phase-independent
index throughout the phases,
having an F-test probability that the improvement is just due to statistical chance
of $2\times10^{-3}$. Using separate power-law indexes
for the three observations
taken at $\phi=0.2$ does not improve the fit. Furthermore, individual fits of the observations
suggest that the photon index is statistically the same among the observations taken at
$\phi=0.2$ and that the photon index at phase 0 is different from that at phase 0.2.
We show the {\it NuSTAR} spectra in Figure~\ref{fig:fig3} and the fit results
in Figures~\ref{fig:fig2} and \ref{fig:fig4}.

For the {\em Swift} data, we extract the spectra using the same regions
as for the timing analysis (Section~\ref{timingana}). 
The center of the source extraction circle was determined for each observation separately.
Since the source spectral properties vary with orbital phase \citep[][]{adk+13},
we performed phase-resolved spectroscopy.
We folded the observations using the new timing solution we found in Section~\ref{timingana}
and merged the data in each phase bin,
for a total of twelve phase bins.
We further produced two spectra for phase 0, one for the high-flux state
and another for the rest of the observations taken at that phase, and eleven spectra
for the other phases hence producing a total of thirteen spectra.
We grouped the data to have 1 count per energy bin because of the paucity
of counts in some phase bins, and used {\it l} statistics.
For the phases that have enough counts, we also tried to fit the spectra
using $\chi^2$ statistics, after grouping to have more than 20 counts per energy bin,
and found that the results
are consistent with those obtained using {\it l} statistics.
We then fit all 13 spectra jointly with an absorbed power-law model with
a common $N_{\rm H}$ (frozen at the {\it XMM-Newton}-measured value
of $7.2\pm0.2\times10^{21}\rm \ cm^{-2}$)
throughout the observations but a separate photon index
and flux for each spectrum. We find that the power-law model
explains the data with photon indices of $\Gamma=1.2-1.8$ and 3--10\,keV fluxes of
$F_{\rm 3-10\ keV}=0.34$--$2.9\times 10^{-12}\rm \ erg\ cm^{-2}\ s^{-1}$, where
the maximum flux was for the high-flux state
(see Figures~\ref{fig:fig2} b and c).

We also tried to determine $N_{\rm H}$ at each orbital phase using the {\it Swift} data.
However, we were able to constrain the fit parameters reasonably only for phase 0
(without the high-flux state).
We fit the spectrum for phase 0 (excluding the high-flux state) with a power-law model, and find
that the best-fit parameters are $N_{\rm H}=7.7\pm1.2\times10^{21}\rm \ cm^{-2}$,
$\Gamma=1.4\pm0.1$, and $F_{\rm 3-10\ keV}=1.5\pm0.1\times 10^{-12}\rm \ erg\ cm^{-2}\ s^{-1}$.
This $N_{\rm H}$ value at phase 0 is consistent with those obtained using the
{\it XMM-Newton} data for phases 0.3 and 0.6 above, suggesting that $N_{\rm H}$ does not
strongly vary as a function of orbital phase. Although
we cannot clearly rule out orbital variation of $N_{\rm H}$,
a $\sim$10\% variation of $N_{\rm H}$ does not significantly change the {\it Swift} results.

Accretion-powered neutron star HMXBs often show spectral features such as
emission lines or an exponential cutoff in the X-ray band \citep[][]{chr+02}.
We find that the
X-ray spectrum of 1FGL~J1018.6$-$5856 is well described with a power-law model
without requiring any additional features (e.g., Figure~\ref{fig:fig3}).
For example, fitting the spectrum for phase 0 with a cutoff power-law model
({\tt pow*highecut} in {\tt XSPEC}) does not improve the fit, and 
the best-fit parameters are not constrained.
We further changed the spectral grouping in order to have the spectra cover
a broader energy range, and to see if a cutoff is required at higher energy.
Specifically, we grouped the {\it NuSTAR} spectra to
have more than 15 counts per energy bin, covering the 3--70\,keV band.
We fit the spectra with a power-law model and a cutoff power-law model, and
found the same results as above; no cutoff is required in the fit.

We performed additional analysis to determine the lower limit
for the cutoff energy
($E_{\rm cutoff}$ of the {\tt highecut} model). However,
it is not possible to set a meaningful lower limit for $E_{\rm cutoff}$ without
constraining the e-folding energy ($E_{\rm f}$ of the {\tt highecut} model).
We therefore limit $E_{\rm f}$ bewteen 6\,keV and 12\,keV, values
obtained for a sample of accretion-powered neutron star HMXBs \citep{chr+02},
and found that the 90\% lower limit for $E_{\rm cutoff}$ is 39\,keV and 34\,keV
for $E_{\rm f}$ of 6\,keV and 12\,keV, respectively.
Note that some accretion-powered black hole binaries are known to have the cutoff
energy above 70\,keV and that our data are not sensitive to such high energy cutoff.

\subsection{Spectral variability}

\begin{figure}
\centering
\hspace{-6mm}
\includegraphics[width=3.6 in]{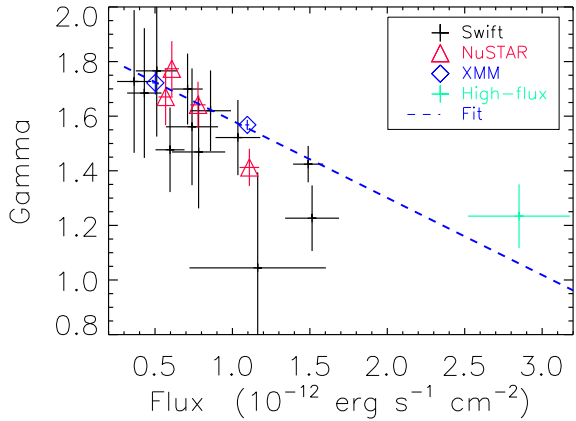} 
\hspace{-4mm}
\vspace{-3mm}
\figcaption{Photon index vs. 3--10\,keV flux. The dashed blue line shows the best-fit
linear function.
\label{fig:fig4}
}
\hspace{1mm}
\vspace{-2mm}
\end{figure}

The spectral hardness varies with orbital phase (Figure~\ref{fig:fig2}c)
and flux (Figure~\ref{fig:fig4}).
Figure~\ref{fig:fig4} shows an apparent correlation
between flux and spectral hardness.
We fit the apparent correlation with a constant function
and found that it does not provide
an acceptable fit ($\chi^2/$dof=55/18).
We therefore added a linear slope to the constant function and
find that the linear fit explains the data well ($\chi^2$/dof=17/17).
The measured slope is $-0.28 \pm 0.04$ (per $10^{-12}\ \rm erg\ cm^{-2}\ s^{-1}$),
consistent with that reported by \citet{adk+13}. The best-fit function is shown in
Figure~\ref{fig:fig4}.

\citet{adk+13} suggested that there is evidence for a correlation between
X-ray flux and spectral hardness. With the new and larger dataset, it is
clear that the two quantities are correlated. For example, we find that the
Spearman's rank order correlation coefficient is $-0.84$ and the significance
is $5.6\sigma$. We further verified that the X-ray flux
and the photon index vary orbitally using the $\chi^2$ test, which resulted in
$p<10^{-10}$ and $p\sim10^{-5}$, respectively. Note that whether or not we include
the high-flux data point in the correlation calculation does not
significantly change the result.

Although the significance for the correlation is high, uncertainties in the
measurements are significant (see Figure~\ref{fig:fig4}) and need to be considered
for the significance calculation.
In order to do so, we performed simulations.
Note that the photon index and the flux are correlated in the spectral fit,
and one needs to take into
account the covariance. We do this by using the covariance matrices in the simulation as
was done by \citet{adk+13}. In 100,000 simulations, a non-negative correlation occurred 316
times, which suggests that the significance of the negative correlation is $\sim$99.7\%.
We also carried out simulations for the linear correlation, and measured
the confidence level of the negative linear correlation to be $\sim$99.9\%.

We also checked for short-term variability ($\sim$10\,ks) using the
longer {\it XMM-Newton} observation (Obs. ID 0694390101) because it has the
best statistics. We calculate the count rate and hardness ratio
(ratio of count rates in two energy bands; e.g, $C_{\rm 3-10\,keV}/C_{\rm 0.3-3\,keV}$)
on various time scales and energy bands. We find variabilities of $\sim$20\% and $\sim$10\%
for the count rate and hardness ratio, respectively, but no correlation between them.

\section{Discussion}
\label{sec:disc}

Our analysis of the {\it NuSTAR}, {\it XMM-Newton}, and {\it Swift} data
is largely consistent with the X-ray results reported by \citet{adk+13},
but the current work provides improvements and refinements.

First, using longer {\it Swift} observations, we find that the orbital period
of 1FGL~J1018.6$-$5856 is $16.544\pm0.008$\,days, consistent with the gamma-ray
measurement \citep[$16.531\pm0.006$\,days;][]{ccc+14}.
When folded on the new period, the light
curve shows two distinct features; a spike at phase 0 and a broad sinusoidal
hump (Figure~\ref{fig:fig2}), similar to those reported previously \citep[][]{fermi12,adk+13}.
With the new period measurement, however, we find that the spike at phase 0 is a
persistent feature and shows less orbit-to-orbit variability
than was suggested by \citet{adk+13}.
Second, we clearly see the correlation between flux and spectral hardness for
which \citet{adk+13} found only marginal evidence.
This is possible thanks to more sensitive observations made with {\it NuSTAR}, {\it Swift} and
{\it XMM-Newton}.

Note that we combined all the {\it Swift} observations taken
over a period of $\sim$2000\,days
for the spectral analysis. If there is long-term ($\gapp P_{\rm orb}$)
and/or short-term (10--100\,ks) variability,
the combined results may be incorrect. This is a concern because
there are only a few observations per orbital phase bin, and
individual exposure of the observations is only $\sim$ks.
Furthermore, if the orbital period is not accurate or varies with time, phases of
later observations will change, introducing an additional error to the analysis.
However, the agreement of the {\it Swift} measurements with the {\it NuSTAR}
and {\it XMM-Newton} results suggests that the errors may not be large compared to the
statistical uncertainties, having no significant impact on the results.
For example, the significance of the correlation between flux and photon index
is still $\gapp$99\% when adding 10\% systematic uncertainty to the {\it Swift}
measurements.

The broadband X-ray spectra of 1FGL~J1018.6$-$5856 at phases 0 and 0.2 are
well described with a power-law model in the 0.5--40 keV band.
Recently, \citet{wr15} find that, based on parameter
space consistent with radial velocity measurements, a neutron star model
is preferred over a typical stellar mass black hole, although both classes are still
allowed for 1FGL~J1018.6$-$5856. We check to see if the source shows any 
evidence for accretion, such as line features or an exponential cutoff
in the X-ray spectrum, as is often seen in neutron star HMXBs,
and find none (Figure~\ref{fig:fig3}).
Furthermore, we find no clear evidence for an exponential
cutoff at $E_{\rm cutoff}<70$\,keV (for spectrum at phase 0). We,
therefore, set the 90\% lower limit
for  $E_{\rm cutoff}$ to be 34--39\,keV for e-folding energies of 6--12\,keV
\citep[$E_{\rm f}$; see][for the range of $E_{\rm f}$ of neutron star HMXBs]{chr+02}.
This lower limit is large for a neutron star HMXB
\citep[typical $E_{\rm cutoff}\sim10-20$\,keV; e.g., see][]{chr+02}.
Note that the X-ray pulsar X~Per (also known as 4U~0352$+$309)
for which \citet{chr+02} did not
find a clear spectral cutoff turned out to have a cutoff at 69\,keV \citep[][]{ltc12},
which is comparable to the energy under which we did not find any evidence for a spectral
cutoff in 1FGL~J1018.6$-$5856. Also, high cutoff energies $\gapp$70\,keV have been seen
in black hole binaries \citep[][]{gjk+99}. Therefore, we cannot clearly rule out the possibility that
1FGL~J1018.6$-$5856 is a black hole binary or a neutron star bianry
with unusually high cutoff energy based only on
the spectral cutoff. Nevertheless, the continuum spectrum of X~Per or
other X-ray binaries is very complex \citep[e.g.,][]{chr+02}
while we see a simple power-law spectrum for 1FGL~J1018.6$-$5856.
This suggests that 1FGL~J1018.6$-$5856 may be a
non-accreting neutron star system, which has also
been suggested for another gamma-ray binary LS~5039 \citep[e.g.,][]{t11}.

In analogy to LS~5039, we may identify the location of
the sinusoidal X-ray peak at
$\phi\sim0.4$ (Figure~\ref{fig:fig2}) as inferior conjunction,
and the gamma-ray peak at $\phi\sim0$ \citep[][]{ccc+14}
as superior conjunction \citep[][]{ktu+09, fermi09}.
Then, the phase difference of $\Delta\phi\sim0.4$
between the two conjunctions implies that the orbit is eccentric.
We note, however, that it is not clear whether the X-ray and gamma-ray
peaks are physically related to the conjunctions or the apastron/periastron
passages, and that
alignment of the X-ray and gamma-ray peaks with inferior and
superior conjunctions may not be precise.
Therefore, more observations and detailed modeling are required in order to
draw a firm conclusion.

We find that the X-ray spectral properties of 1FGL~J1018.6$-$5856
clearly show orbital modulation. Pulsar models for gamma-ray binaries often
attribute such orbital modulation with
orbital variation in the adiabatic cooling timescale \citep[][]{kha+07,kab+08},
the electron injection spectrum, or the location and the shape of the wind
nebula \citep[][]{d06}.
The pulsar models have been applied to the
similar system LS~5039 \citep[e.g., spectral variability
and recurring X-ray flares;][]{ktu+09,adk+13}, and have reproduced the
overall spectral energy distribution \citep[e.g.,][]{d06}.
However, whether or not these models can explain
the spiky feature at phase 0 we see in
1FGL~J1018.6$-$5856 needs to be investigated.

We note that the high-flux state observed with {\it Swift}
at phase 0 (Figure~\ref{fig:fig2}a and b)
is not reproduced in other observations taken at the same phase. It may be because the
two observations in the high-flux state were made in a very narrow phase interval and the
later observations did not cover that phase interval.
In order to see if this is the case, we first verified that the high-flux state was
not produced by short timescale variability ($\sim$ks);
it lasted for the full duration of the exposures of the observations
(24 ks at MJD~55585.7 and 7 ks at MJD~55618.7
for Obs. IDs 00031912004 and 00031912011, respectively) which cover
a phase interval of $\Delta \phi=0.022$ (at $\phi=0.034\pm0.011$ for $P_{\rm orb}=16.54$\,days).
We then measured the phases of the other observations. We find that
there are four observations made at the high-flux phase interval,
and none of them was in the high-flux state. Since the phase of an observation
can change significantly for a different orbital period, we further varied $P_{\rm orb}$ within
the measurement uncertainty of $0.01$\,day, and find the same result. This suggests
that there is orbit-to-orbit flux variability at phase 0.

We find that the duration of the high-flux state is longer than 24\,ks and shorter
than 1.8\,days. The minimum duration is set to be 24\,ks because the high-flux states last during
the observation (see above). The maximum duration is set to be the interval between a
high-flux state and the next non-high-flux observation, which is 1.8\,days for both
high-flux states. As noted by \citet{adk+13}, the observational properties of the flare
such as duration and orbital repeatability look more like that of
LS~5039 \citep[e.g.,][]{ktu+09}
than those of LS~I~$+$61$^\circ$303 \citep[e.g.,][]{ltz+11,skh+09}.
This may support the idea that the flares are produced by clumpiness
of the stellar wind \citep[e.g.,][]{znc10} since the stellar companion
(Be star) of LS~I~$+$61$^\circ$303 is different from those (O stars)
of 1FGL~J1018.6$-$5856 and LS~5039 as the flare properties do.
However, how the clumpiness produces flares at one orbital phase
for 1FGL~J1018.6$-$5856 or LS~5039 but not at random orbital phases needs
to be further investigated.

\section{Conclusions}
\label{sec:concl}
We present results of {\it NuSTAR}, {\em Swift}, and {\em XMM-Newton} observations
of the gamma-ray binary 1FGL~J1018.6$-$5856.
Using the {\it Swift} data, we measured the orbital period of the source to be 16.544$\pm$0.008 days,
in agreement with the refined gamma-ray measurement of \citet{ccc+14}.
The new period is only slightly
different from that used in our previous X-ray study, and hence our spectral and temporal
analysis results agree well with the previous X-ray measurements.
We find that the flux enhancement at phase 0 occurs more regularly in time than was
suggested previously based on {\it Swift} data. The new {\it NuSTAR} and {\it XMM-Newton}
data allow us to show clearly the correlation between X-ray flux and spectral hardness of
1FGL~J1018.6$-$5856. Finally, the broadband X-ray spectrum of 1FGL~J1018.6$-$5856 suggests
that it may not be an accretion-powered system.\\

We thank R.~W.~Romani for useful discussions.
This work was supported under NASA Contract No. NNG08FD60C,
and  made use of data from the {\it NuSTAR} mission,
a project led by  the California Institute of Technology, managed by the Jet Propulsion  Laboratory,
and funded by the National Aeronautics and Space  Administration. We thank the {\it NuSTAR} Operations,
Software and  Calibration teams for support with the execution and analysis of  these observations.
This research has made use of the {\it NuSTAR}  Data Analysis Software (NuSTARDAS) jointly developed by
the ASI  Science Data Center (ASDC, Italy) and the California Institute of  Technology (USA).
This research has made use of data obtained from
the High Energy Astrophysics Science Archive Research Center
(HEASARC), provided by NASA's Goddard Space Flight Center.
H.A. acknowledges supports provided by the NASA sponsored Fermi Contract NAS5-00147 and by
Kavli Institute for Particle Astrophysics and Cosmology (KIPAC).
LN wishes to acknowledge the Italian Space Agency (ASI) for financial
support by ASI/INAF grant I/037/12/0-011/13.


\begin{thebibliography}{48}
\expandafter\ifx\csname natexlab\endcsname\relax\def\natexlab#1{#1}\fi

\bibitem[{Abdo} {et~al.}(2009){Abdo}]{fermi09}
{Abdo}, A.~A., {Ackermann}, M., {Ajello}, M., et~al. 2009a, \apj, 706, L56

\bibitem[{Abramowski} {et~al.}(2012){Abramowski}]{hess12}
{Abramowski}, A., {Acero}, F., {Aharonian}, F. et~al. 2012, \aap, 541, A5

\bibitem[{Ackermann} {et~al.}(2012){Ackermann}]{fermi12}
{Ackermann}, M., {Ajello}, M., {Ballet}, J., et~al. 2012, Science, 335, 189

\bibitem[{An} {et~al.}(2013){An}]{adk+13}
{An}, H., {Dufour}, F., {Kaspi}, V.~M., \& {Harrison}, F.~A. 2013, \apj, 775, 135

\bibitem[{Anders} \& {Grevesse} (1989){Anders} \& {Grevesse}]{ag89}
{Anders}, E., \& {Grevesse}, N. 1989, \gca, 53, 197

\bibitem[{Bogovalov} {et~al.} (2008){Bogovalov}, {Khangulyan}, {Kolodoba}, {Ustyugova}, \& {Aharonian}]{bkk+08}
{Bogovalov}, S.~V., {Khangulyan}, D.~V., {Kolodoba}, A.~V., {Ustyugova}, G.~V., \& {Aharonian}, F.~A. 2008,
\mnras, 387, 63

\bibitem[{Bosch-Ramon} \& {Paredes}(2004){Bosch-Ramon}, \& {Paredes}]{bp04}
{Bosch-Ramon}, V., \& {Paredes}, J.~M. 2004, \aap , 417, 1075

\bibitem[{Bosch-Ramon} \& {Khangulyan}(2009){Bosch-Ramon}, \& {Khangulyan}]{bk09}
{Bosch-Ramon}, V., \& {Khangulyan}, D. 2009, Int. J. Mod. Phys. D, 18, 347

\bibitem[{{Buccheri} {et~al.}(1983){Buccheri}, {Bennett}, {Bignami}, {Bloemen},
{Boriakoff}, {Caraveo}, {Hermsen}, {Kanbach}, {Manchester}, {Masnou}, {Mayer-Hasselwander},
{\"Ozel}, {Paul}, {Sacco}, {Scarsi} \& {Strong}}]{bbb+83}
{Buccheri}, R., {Bennett}, K., {Bignami}, et~al. 1983, \aap, 128, 245

\bibitem[{Capalbi} {et~al.}(2005){Capalbi}]{cps+05}
{Capalbi}, M., {Perri}, M., {Saija}, B., {Tamburelli}, F., \& {Angelini}, L. 2005,
The Swift XRT Data Reduction Guide, Technical Report 1.2

\bibitem[{Chernyakova} {et~al.} (2006){Chernyakova}, {Neronov}, \& {Walter}]{chw06}
{Chernyakova}, M., {Neronov}, A., \& {Walter}, R. 2006, \mnras, 372, 1585

\bibitem[{Coley} {et~al.} (2014){Coley}]{ccc+14}
{Coley}, J.~B., {Corbet}, R.~H.~D., {Cheung}, C.~C., {Coe}, M.,
{Edwards}, P., {McBride}, V., {McSwain}, M.~V., {Stevens}, J. 2014, 14, 122.10,
http://adsabs.harvard.edu/abs/2014HEAD...1412210C

\bibitem[{Coburn} {et~al.} (2002){Coburn}]{chr+02}
{Coburn}, W., {Heindl}, W.~A., {Rothschild}, R.~E.,
{Gruber}, D.~E., {Kreykenbohm}, I., {Wilms}, J., {Kretschmar}, P.,
{Staubert}, R., 2002, \apj, 580, 394

\bibitem[{Dubus} (2006){Dubus}]{d06}
{Dubus}, G. 2006, \aap, 456, 801

\bibitem[{Dubus} (2013){Dubus}]{d13}
{Dubus}, G. 2013, \aapr, 21, 64

\bibitem[{Dubus} {et~al.}(2010){Dubus}, {Cerutti}, \& {Henri}]{dch10}
{Dubus}, G., {Cerutti}, B., \& {Henri}, G. 2010, \mnras, 404, L55

\bibitem[{{Grove} {et~al.}(1999){Grove}}]{gjk+99}
{Grove}, J.~E., {Johnson}, W.~N., {Kroeger}, R.~A.,
{McNaron-Brown}, K., {Skibo}, J.~G., {Phlips}, B.~F. 1999, \apj, 500, 899

\bibitem[{{Harrison} {et~al.}(2013){Harrison}}]{hcc+13}
{Harrison}, F.~A., {Craig}, W.~W., {Christensen}, F.~E., et~al. 2013, \apj, 770, 103

\bibitem[{Johnston} {et~al.}(1992){Johnston}]{jml+92}
{Johnston}, S., {Manchester}, R.~N., {Lyne}, A.~G., et~al. 1992, \apj, 387, L37

\bibitem[{Kaspi} {et~al.}(1995){kaspi}]{ktn+95}
{Kaspi}, V.~M., {Tavani}, M., {Nagase}, F., et~al. 1995, \apj, 453, 424

\bibitem[{Kaufman-Bernad{\'o}} {et~al.}(2002){Kaufman-Bernad{\'o}}, {Romero}, \& {Mirabel}]{krm02}
{Kaufman-Bernad{\'o}}, M.~M., {Romero}, G.~E., \& {Mirabel}, I.~F. 2002, \aap, 385, L10

\bibitem[{Khangulyan} {et~al.}(2007){Khangulyan}]{kha+07}
{Khangulyan}, D., {Hnatic}, S., {Aharonian}, F., \&
{Bogovalov}, S., 2007, \mnras, 380, 320

\bibitem[{Khangulyan} {et~al.}(2008){Khangulyan}]{kab+08}
{Khangulyan}, D.~V., {Aharonian}, F.~A., {Bogovalov}, S.~V.,
{Koldoba}, A.~V., \& {Ustyugova}, G.~V., 2008, Int. J. Mod. Phys. D, 17, 1909

\bibitem[{Kishishita} {et~al.}(2009){Kishishita}, {Tanaka}, {Uchiyama} \& {Takahashi}]{ktu+09}
{Kishishita}, T., {Tanaka}, T., {Uchiyama}, Y., \& {Takahashi}, T. 2009, \apj, 697, L1

\bibitem[{Leahy} (1987) {Leahy}]{l87}
{Leahy}, D.~A. 1987, \aap, 180, 275

\bibitem[{{Li} {et~al.}(2011a) {Li}, {Torres}, {Chen}, {G{\"o}tz}, {Rea}, {Zhang}, {Caliandro} \&
{Wang}}]{ltc+11} {Li}, J., {Torres}, D.~F., {Chen}, Y., et~al. 2011, \apj, 738, L31

\bibitem[{{Li} {et~al.}(2011b) {Li}, {Torres}, {Zhang}, {Chen}, {Hadasch}, {Ray}, {Kretschmar}, {Rea} \&
{Wang}}]{ltz+11} {Li}, J., {Torres}, D.~F., {Zhang}, S., et~al. 2011, \apj, 733, 89

\bibitem[Loredo (1992)]{l92}
Loredo, T.~J. 1992, in Statistical Challenges in Modern Astronomy, ed. E.~D. Feigelson \&
G.~J. Babu (New York: Springer), 275

\bibitem[{{Lutovinov} {et~al.}(2012) {Lutovinov}, {Tsygankov}, \&
{Chernyakova}}]{ltc12} {Lutovinov}, A., {Tsygankov}, S., \& {Chernyakova}, M., 2012, \mnras, 423, 1978

\bibitem[{Mirabel} (2012) {Mirabel}]{m12}
{Mirabel}, I.~F. 2012, Science, 335, 13

\bibitem[{Romero} {et~al.}(2003){Romero}, {Torres}, {Kaufman-Bernad{\'o}}
\& {Mirabel}]{rtk+03}
{Romero}, G.~E., {Torres}, D.~F., {Kaufman-Bernad{\'o}}, M.~M. \& {Mirabel}, I.~F. 2003, \aap, 410, L1

\bibitem[{Sierpowska-Bartoski} \& {Torres} (2008){Sierpowska-Bartoski} \& {Torres}]{st08}
{Sierpowska-Bartoski}, A., \& {Torres}, D.~F. 2008, Astropart. Phys., 20, 239

\bibitem[{Smith} {et~al.}(2009)]{skh+09}
{Smith}, A., {Kaaret}, P., {Holder}, J., {Falcone}, A.,
{Maier}, G., {Pandel}, D., {Stroh}, M. 2009, \apj, 693, 1621

\bibitem[{Takahashi} {et~al.}(2009)]{tku+09}
{Takahashi}, T., {Kishishita}, T., {Uchiyama}, Y., et~al. 2009, \apj, 697, 592

\bibitem[{Takata} {et~al.} (2012){Takata}, {Okazaki}, {Nagataki}, {Naito}, {Kawachi}, {Lee},
{Mori}, {Hayasaki}, {Yamaguchi} \& {Owocki}]{ton+12}
{Takata}, J., {Okazaki}, A.~T., {Nagataki}, S., et~al. 2012, \apj, 750, 70

\bibitem[{Tavani} {et~al.} (1994){Tavani}, {Arons}, \& {Kaspi}]{tak94}
{Tavani}, M., {Arons}, J., \& {Kaspi}, V.~M. 1994, \apj, 433, L37

\bibitem[{Tavani} \& {Arons} (1997){Tavani}, \& {Arons}]{ta97}
{Tavani}, M., \& {Arons}, J. 1997, \apj, 477, 439

\bibitem[{Torres}(2011){Torres}]{t11}
{Torres}, D.~F. 2011, in Proc. First Session of the Sant Cugat Forum on Astrophysics, High-Energy
Emission from Pulsars and their Systems, ed. N. Rea \& D.~F. Torres (Berlin:Springer), 532

\bibitem[{Waisberg} \& {Romani} (2015){Waisberg}, \& {Romani}]{wr15}
{Waisberg}, I., \& {Romani}, R.~W. 2015, \apj, in press, arxiv:astro-ph/1504.05125

\bibitem[{Willms}, {Allen} \& {McCray} (2000){Willms}, {Allen} \& {McCray}]{wam00}
{Willms}, J., {Allen}, A., \& {McCray}, R. 2000, \apj, 542, 914

\bibitem[{Zdziarski}, {Neronov} \& {Chernyakova} (2010)]{znc10}
{Zdziarski}, A.~A., {Neronov}, A., \& {Chernyakova}, M. 2010, \mnras, 403, 1873

\end{thebibliography}
\end{document}